# Exploring Conceptual Modeling Metaphysics: Existence Containers, Leibniz's Monads and Avicenna's Essence

Sabah Al-Fedaghi*

*Computer Engineering Department*
*Kuwait University*
*Kuwait*

salfedaghi@yahoo.com, sabah.alfedaghi@ku.edu.kw

*Abstract* – **Requirement specifications in software engineering involve developing a conceptual model of a target domain. The model is based on ontological exploration of things in reality. Many things in such a process closely tie to problems in metaphysics, the field of inquiry of what reality fundamentally is. According to some researchers, metaphysicians are trying to develop an account of the world that properly conceptualizes the way it is, and software design is similar. Notions such as classes, object orientation, properties, instantiation, algorithms, etc. are metaphysical concepts developed many years ago. Exploring the metaphysics of such notions aims to establish quality assurance though some objective foundation not subject to misapprehensions and conventions. Much metaphysical work might best be understood as a model-building process. Here, a model is viewed as a hypothetical structure that we describe and investigate to understand more complex, real-world systems. The purpose of this paper is to enhance understanding of the metaphysical origins of conceptual modeling as exemplified by a specific proposed high-level model called thinging machines (TMs). The focus is on thimacs (*thi*ngs/*mac*hine) as a single category of TM modeling in the context of a two-phase world of staticity and dynamics. The general idea of this reality has been inspired by Deleuze's 'the virtual' and related to the classical notions of Leibniz's monads and Avicenna's essence. The analysis of TMs leads to several interesting results about a thimac's nature at the static and existence levels.**

*Index Terms* – **Conceptual model, software notions, ontology, metaphysics, software requirement development.**

## I. INTRODUCTION

Every science presupposes and requires some sort of metaphysical foundation [1][2]. Metaphysics can be defined as the study of categories with the purpose of identifying them, defining them if possible, and determining the relationships among these categories [3]. According to [4],

> Metaphysics, at bottom, is about the fundamental structure of reality. Not about what's necessarily true. Not about what properties are essential. Not about conceptual analysis. Not about what there is. Inquiry into necessity, essence, concepts, or ontology might help to illuminate reality's structure. But the ultimate goal is insight into this structure itself—insight into what the world is like, at the most fundamental level. Despite a long tradition of challenges to the viability of the metaphysical enterprise, metaphysics is once again a thriving subdiscipline within philosophy [5].

According to [6], metaphysics involves the construction and evaluation of *model* classes. A model is an imagined or hypothetical structure that we describe and investigate to understand some more complex, real-world target system or domain. A growing number of scientists in the modeling community are exploring ways to address issues related to ontology. Some philosophers recognize the value of models for providing "a revolution in metaphysics issues" [7]. Reference [5] stated that much metaphysical work, especially of the contemporary systematic kind, might best be understood as *model-building*, in a specific sense of this term that draws on recent philosophy of science.

Specially, [7] claims that developments in "*computer modeling* […]" have the potential to contribute to what may be the most significant change in Western philosophy since the foundational work of Aristotle's teacher Plato in the 4th century BCE." In computer science, finding solutions to practical problems employs models of the world and applied metaphysics [8]. In this paper, the focus is on *conceptual modeling* utilized to describe proposed requirements and designs for software systems. Conceptual models are typically represented in terms of a graphic structure.

Metaphysics is very close to *conceptual modeling*. Notions such as classes, object orientation, properties, instantiation, algorithm, etc. are metaphysical concepts developed thousands of years ago. Exploring the metaphysics of such notions aims to establish quality assurance though some objective foundation not subject to misapprehensions and conventions. The closeness of metaphysics and conceptual modeling has driven some researchers to propose the feasibility of the graphical representation of philosophical concepts in terms of UML (unified modeling language) diagrams and methods [9]. According to [9], this approach could help philosophy in mapping, explaining, clarifying, and model-checking because UML has the capability of representing abstract conceptual structures in a highly standardized and formalized manner.

Conceptual modeling in *software engineering* concerns metaphysics. An example is viewing the software as a unique sort of *being*. According to [10], researchers have not yet captured what software actually is or how software can be characterized and need to understand the ontological status of software as the product of software engineering. Reference [11] viewed the notion of algorithm, which is, metaphorically speaking, "the 'soul' of a computer program," in eternal forms [10]. Some software engineering researchers have already contributed to understanding similar issues, such as programming and software system design. According to [12], "Metaphysicians are aiming for theories that describe the fundamental structure, nature, or entities in the world. They're trying to develop an account of the world that properly conceptualizes the way it is. Software design is the same."

The new contribution of this paper consists of a metaphysical exploration of conceptual modeling, which is one of the central activities in software engineering, as an intermediate artifact for system construction. Specifically, such an exploration is applied to develop a new high-level model called thinging machine (TM) modeling [13]-[15]. Focusing on the TM is motivated by the subtle nature of TMs that include *one*-category ontology (things/machines) and *five* actions (create, process, release, transfer and receive). This facilitates easier mapping to metaphysical notions. The results contribute to further understanding TM modeling, in addition to introducing some metaphysical insights of conceptual modeling.

### A. Aims

In the field of conceptual modeling, ontologies provide a foundational theory to describe the structure and behavior of the modeled domain (e.g., [1]). Such ontology-driven conceptual modeling is defined as the utilization of ontological theories to develop engineering artifacts to improve the theory and practice of modeling [16]. One purpose of such an approach is to improve the process and quality of engineering software systems because ontologies provide real-world semantics for language constructs or assess the adequacy and sufficiency of modeling constructs that represent constructs in domains [16]. According to [16], in this context, several research gaps and shortcomings can be identified that still pose challenges for further development in the field (e.g., model comprehension, complexity). More research would be beneficial for the field of ontology-driven conceptual modeling, as the principal purpose of a model is to be understood and comprehended by anyone who uses it [16].

This paper aims at gaining insight into the model of TMs. Specifically; the paper concentrates on the metaphysical characteristics of TMs by analyzing its notions, such as event, time, object, action, etc. Metaphysics here is viewed as an interpretation of TM modeling in terms of creation, processing, releasing, transferring and receiving for the sphere of actions common to all TM-proposed universe of elements: thimacs (thing/machines).

The paper is a continuation of sequel of research work about TMs. For example, TM modeling describes the dual status of reality that involves studying the questions in Fig. 1.

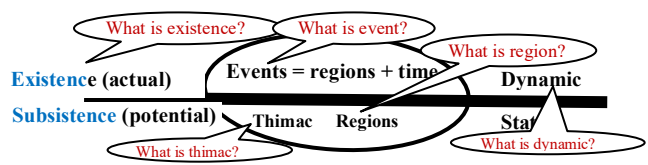

Fig. 1 Some metaphysics question in the context of TM

### B. Content of paper

The next section includes an enhanced description of the TM model as a world of *thimacs* – a network of thimacs that articulate the underlying structure of the world. Subsections are devoted to what is a TM thing and what is a TM machine? Section 3 illustrates TM modeling with two examples:
- The first example involves modeling the biography of a fictional man (taken from [17]-[19]).
- The second example models a chain-store company offering various types of cheese.

Previous papers about TMs give many examples; however, for diversity purpose, the two examples above are different because their focus is on table representations with time-based data.

Section 4 explores thimacs at two levels of potentiality/actuality, adopting an idea that goes back to the Stoic view of reality. Things subsist in the static level (plane of reality) and are brought above the threshold of subsistence/existence to actuality. In Section 5, the interiority of the thimac is analyzed in terms of two directions: the thimac acting on itself and the thimac as an agent that acts on other things (objects).

This analysis leads to the interesting results of defining *what TM existence is*, as illustrated in Fig. 2.

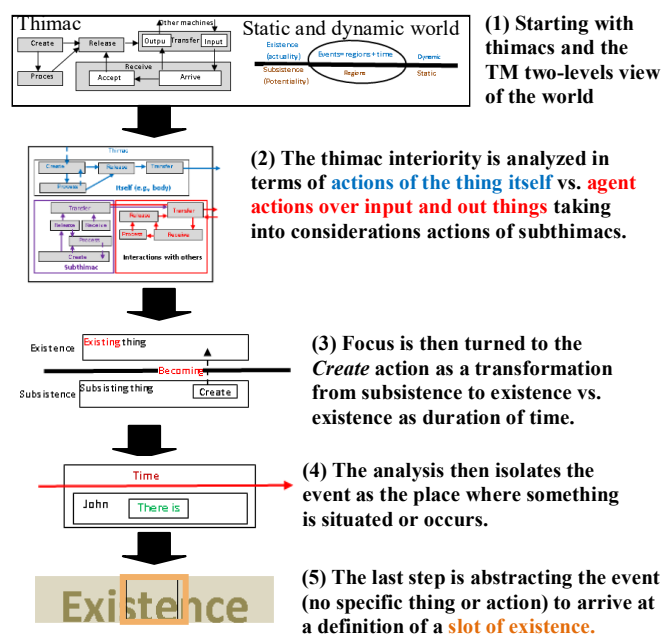

Fig. 2 Summary of paper content

## II.    THE TM MODEL

The TM diagrammatic model has been applied in many applications. Each TM paper presents an application example; furthermore, each paper introduces additional development of the model as an underlying structure of the world. This paper focuses on *the metaphysical TM aspects*.

This TM modeling approach may relate to the view that at some fundamental level, the world is a graph of nodes and edges. The concrete world is a single, large structure best analyzed as a diagram [20]. According to [21], dissatisfaction with "the resources of set theoretic semantics of predicate logic for describing ontological structure" motivates selecting graph theory, aiming to show that we can conceive of the underlying structure of the world as a graph." A physical individual is a collection of joined subgraphs, each of which is a temporal part of the individual. Physical individuals come into and go out of existence, and hence, the necessity of the existence of the nodes does not entail the necessity of physical individuals [20].

In TM modeling, the response to the question, "what is there?" is a world of *thimacs* – a network of thimacs that articulate the underlying structure of the world. Thimacs are the basic wholes (entities and processes) of composition that include thimacs that can be divided into subthimacs, which can in turn be divided, and so on. Every thimac is distinct from every other thimac by its superthimacs or subthimacs. Additionally, thimacs may attach themselves to other thimacs and form new thimacs.

Each thimac is woven from and in other thimacs, forming an organized whole. According to such a view, the whole holds together as one thimac. The whole is more than a mere entanglement of interconnections of similar thimacs. The thimac as a machine is the basic unit of the whole constructed as a repetition of crystal lattice structure.

Thimacs are the only foundational elements of regions (subdiagrams that represent atemporal things we *talk* about) and corresponding "existence pieces" (events) in reality. Reality is viewed as a composite thimac with a singular unified totality. Hereafter, a thimac may be referred to as a *thing* or *machine* (see Fig. 3).

A thimac is a *machine* when it acts on other thimacs, and it is a *thing* when it is the object of actions by other thimacs. A machine *things* (*Heideggerian* verb); that is, it creates, processes (changes), receives, transfers and releases. Thimacs are things that subsist of their own accord at the static level. They come into existence, persisting through time, and change when they manifest at the dynamic level. Being at a specific position is a standing in certain relations (inside/outside, flow connection) with other thimacs.

### A.    The Machine

According to [22], "To be is to be a machine. Rocks are machines, stars are machines, trees are machines, people are machines, corporations are machines, revolutionary groups are machines, tardigrades are machines." In TM modeling, the thimac *machine* executes five actions: *create*, *process*, *release*, *transfer* and *receive* (see Fig. 4). Each of these static (outside time) actions is a capacity or power to act and becomes a *generic event* when merged with time. Thimacs are realized by creating, processing, releasing, transferring and/or receiving thimacs. Each thimac is affected by/affects the thimacs in contact with it through releasing, transferring and receiving.

A thimac's actions, in Fig. 4, are described as follows.
1) *Arrive:* A thing arrives to a machine.
2) *Accept:* A thing enters the machine. For simplification, the arriving things are assumed to be *accepted* (see Fig. 4); therefore, *arrive* and *accept* combine actions into the *receive* action.
3) *Release:* A thing is ready for transfer outside the machine.
4) *Process:* A thing is changed, handled and examined, but no new thing is generated.
5) *Transfer:* A thing is input into or output from a machine.
6) *Create:* A new thing manifests in a machine.

### B.    The thing

With a thing as a thimac, a *thing* is what can be created, processed, released, transferred and/or received. For example, the sentence *things take time* denotes what are "*being created (brought into existence), processed, released, transferred and/or received*" take time. Each of these things may be composed of (sub)things. In TMs, what is created is established or substantiated as a thing. A TM's *existence* designates manifestation in time and *subsistence* (discussed later), indicating timeless establishment, e.g., essence (discussed later). Additionally, the thimac exists and subsists as a machine that creates, processes, releases, transfers and/or receives.

A thimac could be a composite entity of networks of subthimacs that are a *unified whole*. This thimac will be a *megathimac* or simply a *megamac* (mega machine).

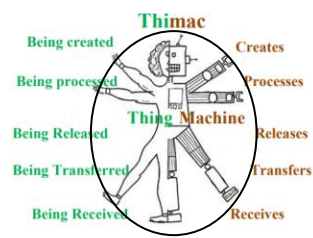

**Fig. 3 The thimac as a thing and a machine (From [23])**

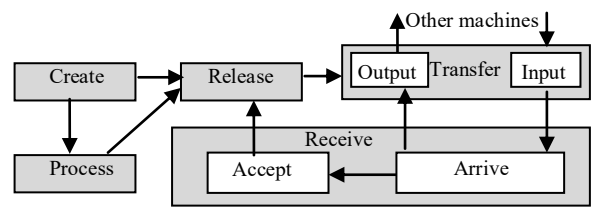

**Fig. 4 Thinging machine without process**

For example, an immaterial thimac, such as *traffic*, is a megamac, as traffic is a complex of machines manifested (exists and subsists) by processing cars, roads, lights, rules, etc. These thimacs organize into a whole TM region such that they are responsible for the thimac of traffic. Note that a megamac is not an arbitrary composite or aggregate of a thimac, e.g., like human-plus-stone, but it has some kind of acting *unity*, e.g., transfer, receive, process and release as a self-unit. Weather, task, transportation and so forth, each is what creates, processes, releases, transfers and receives.

At the dynamic level, this megamac as a collection of related events is a *megaevent*. An example of a unified whole as megaevent is the Aristotelian thesis that *action* and *passion* form a certain kind of whole [24].

In TMs, every existing thing is an event. We – and everything that is – are events [25] (referencing Alfred North Whitehead), we are, "inhaling air, ingesting food, absorbing heat or cold, sweating, defecating, shedding hair and skin. On atomic, molecular, biochemical, cellular, biosystemic, bodily, even conscious levels, we are not stable substances at all. We are constantly engaging in a give-and-take with the rest of creation, all simultaneously" [25].

### III.   EXAMPLES

This section illustrates TM modeling with two examples:
- The first example involves modeling the biography of a fictional man.
- The second example models a chain-store company, offering various types of cheese.

The two examples focus on table representations with time-based data.

#### A.   Modeling the life of John Doe

This example involves the following short biography of a fictional man (taken from [17]-[19]).

John Doe was born on April 3, 1975, as the son of Jack Doe and Jane Doe, who lived in Smallville. Jack Doe registered the birth of his first-born on April 4, 1975, at the Smallville City Hall. John grew up and graduated in 1993. After graduation, he went to live in Bigtown. Although he moved out on August 26, 1994, he forgot to register the change of address officially. It was only at the turn of the seasons that he did a few days later on December 27, 1994. John accidentally died on April 1, 2001.

To store the life of John Doe in a database, a table, Fig. 5, is used to list the time of events in his life [17]-[19]. The method of analyzing the events is based on the natural language description. An alternative analysis on TM modeling is shown in Fig. 6. The diagram represents the biography of John as a whole megamac. Thimacs: *person*, *address 1*, *address 2*, *father* and *database* indicate the entities that exist at the TM events level. The action *create* in this picture denotes *potential* existence over the time period of the biography. Note that for simplification, sometimes *create* is not included under the assumption that the rectangle implicitly implies that. Fig. 6 is a picture of a static timeless world — note the passive voice in the description. Dashed arrows in the figure denote *triggering*.

Thus, in Fig. 6, a person is born (number 1 in the figure) at address 1 (2). His father (3) goes to the database system (4) and registers him (5). The person is processed (graduation process, 6) then is released and transferred to address 2 (7). There, he registers his new address (8 and 9). Eventually, he dies 10.

The static model in Fig. 6 is atemporal and the sequence of actions (e.g., moving from address 1 to address 2) express a *logical* order. The corresponding events model (see Fig. 7) can be described (depending on how events are cut) as follows.

$E_1$: John is born at address 1.
$E_2$: Father reports John's birth.

| Date | |
|---|---|
| April 3, 1975 | John is born |
| April 4, 1975 | John's father officially reports John's birth |
| August 26, 1994 | After graduation, John moves to Bigtown, but forgets to register his new address |
| December 26, 1994 | Nothing |
| December 27, 1994 | John registers his new address |
| April 1, 2001 | John dies |

Fig. 5 Events in the life of John Doe

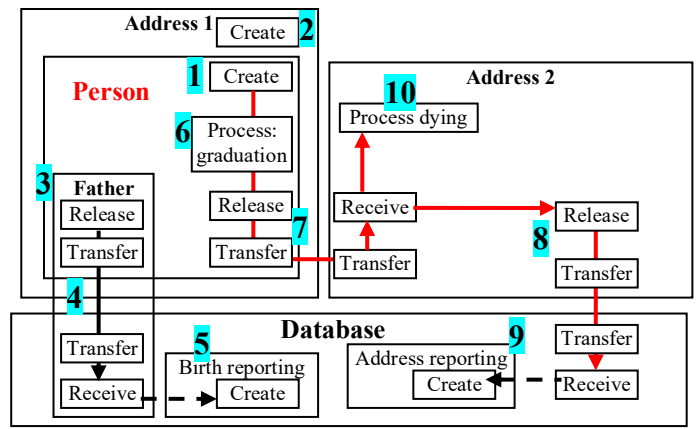

Fig. 6 Static Model of the life of John Doe

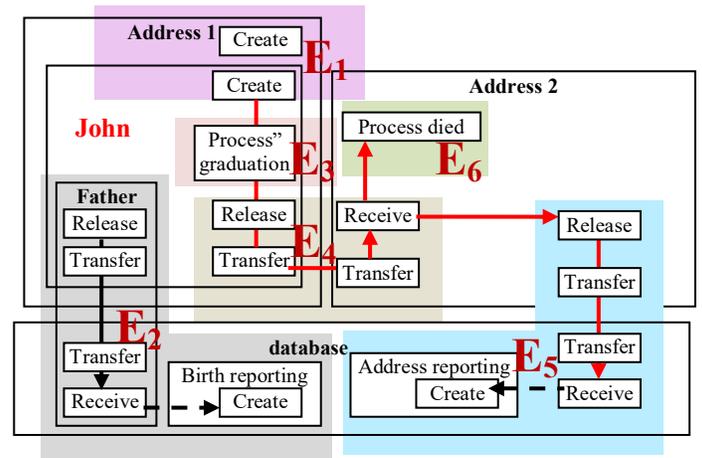

Fig. 7 Events Model (megaevent) of the life of John Doe

$E_3$: John's graduation

$E_4$: John moves to a new address.

$E_5$: John registers the new address.

$E_6$: John dies.

Note that each event by definition includes its time and other descriptions, such as duration and who reports the event. An event is defined as a subdiagram of the static model (region of the event) plus time. Fig. 7 is a *megaevent* that would form the foundation for any type of implementation in terms of tables discussed in such sources as [17]-[19]. The TM model (e.g., Fig. 6) is a unified whole (a single state of affairs) forced by complementary subthimacs.

Fig. 8 shows the chronology of events. If we want to represent the example in terms of events, then the database is a list of events, and each event contains its time stamp.

Contrasting the English and tables technique of [17]-[19] vs. TM modeling shows that the former method is based on an English description with a very large vocabulary, whereas the TM model uses just five basic verbs in addition to identifying entities (person, father, address, etc.). The TM model follows a systematic development in terms of a static description and dynamic specification and then identifies the chronology of events. This paper aims to clarify the metaphysical notions in TMs such as event, existence, etc. that give TM modeling additional advantages of comparison with the ambiguity of natural language.

### B. Temporal databases

Consider the example given by [26] of a company called *CheeseHut* that is a chain store offering three types of cheese, namely young, mature and old cheese, with each cheese having its own price (see Fig. 9). Fig. 10 shows the product history table of CheeseHut as given in [26].

In TM, the modeling involves information about cheese in the form of a table. The table can be viewed as a thimac and its instantiation is a megaevent. Thus, the table has subthimacs and actions.

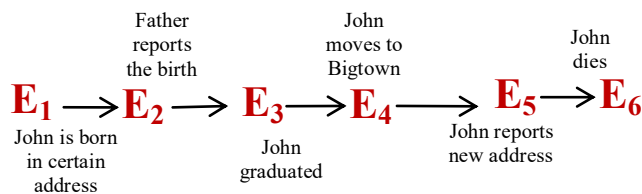

**Fig. 8 Chronology of events in the life of John.**

Accordingly, the static description of a table is shown in Fig. 11. First the table is declared (is created) – see number **1**, and at this holistic level (the table itself). The table, at this holistic level, has a subthimac (**2**) that keeps tracking and updating the number of rows in the table. For simplicity, the mechanism for deleting a row is not shown. The table also has the subthimac row (**3**) that may have its own global information row, such as the number of attributes (not included in the figure). Then, there is the body of the row (**4**) which is constituted by the row key (**5**), ID (**6**), name (**7**) and price (**8**) of the product. These values are constructed though an outside input (**9**) that causes, also, the number of rows in the table to update (**10**). Fig. 12 gives an example of inserting a tuple in the table. An initial assumption is that the table has no tuples at the time of creation.

| Row | id | name | price |
|-----|-----|-------|-------|
| 1 | 1 | Young | 6 |
| 2 | 2 | Mature | 8 |
| 3 | 3 | Old | 12 |

**Fig. 9 Product history table of CheeseHut**

| Row | id | startTime | endTime | name | price |
|-----|-----|-----------|---------|------|-------|
| 1 | 1 | 2011-01-01 | 9999-12-31 | Young | 6 |
| 2 | 2 | 2011-01-01 | 9999-12-31 | Mature | 8 |
| 3 | 3 | 2011-01-01 | 2014-01-01 | Old | 11 |
| 4 | 3 | 2014-01-01 | 9999-12-31 | Old | 12 |

**Fig. 10 Product history table of CheeseHut**

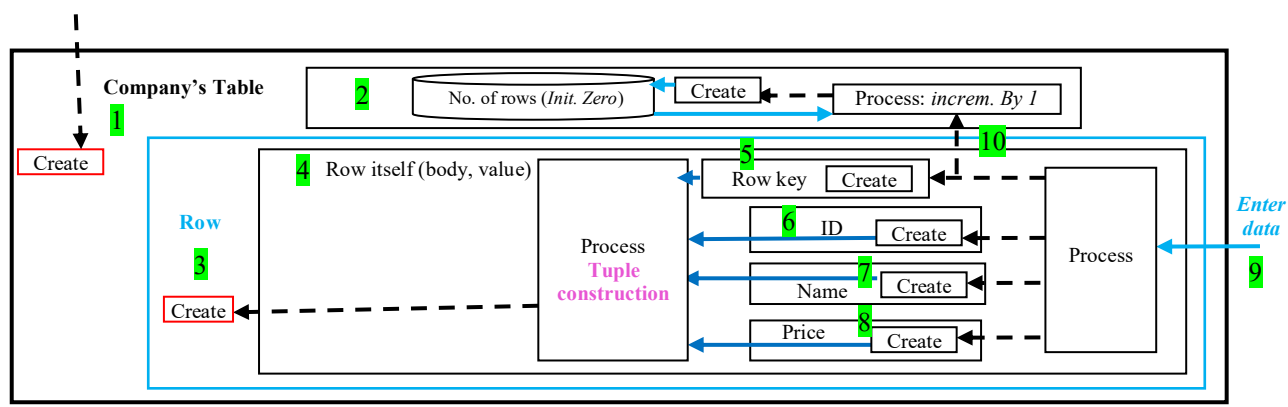

**Fig. 11 The static description of a table**

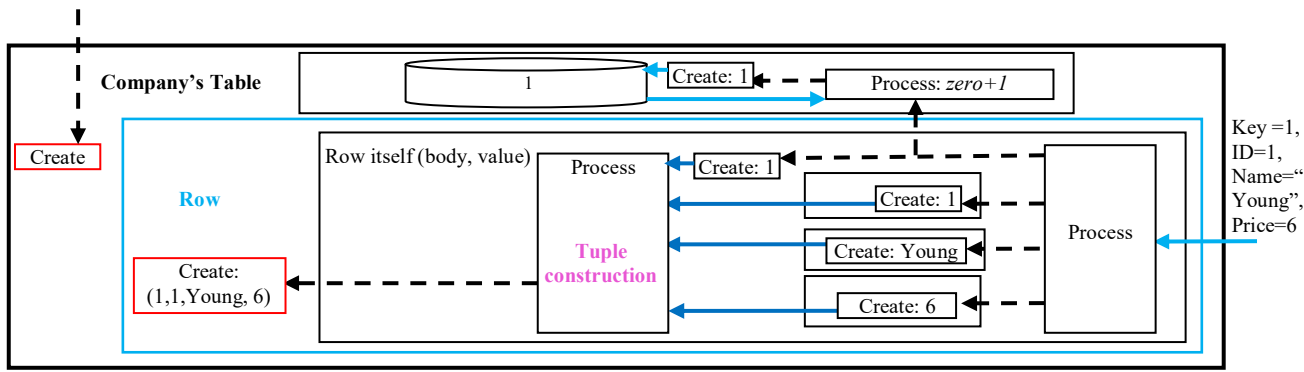

**Fig. 12 Example: Inserting a tuple in the table**

Then, [26] utilizes the history table of CheeseHut when in 2014, CheeseHut changes the supplier for the old cheese, which results in a higher price of 12 euros per kilogram. Note that this history table is a table of TM events. Fig. 13 shows the TM events model, and Fig. 14 shows its corresponding chronology of events. To save space, these figures will not be described further.

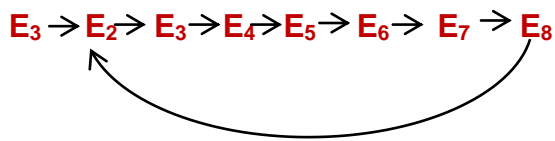

**Fig. 14 Events of creating a row and adding it to a table, repeatedly**

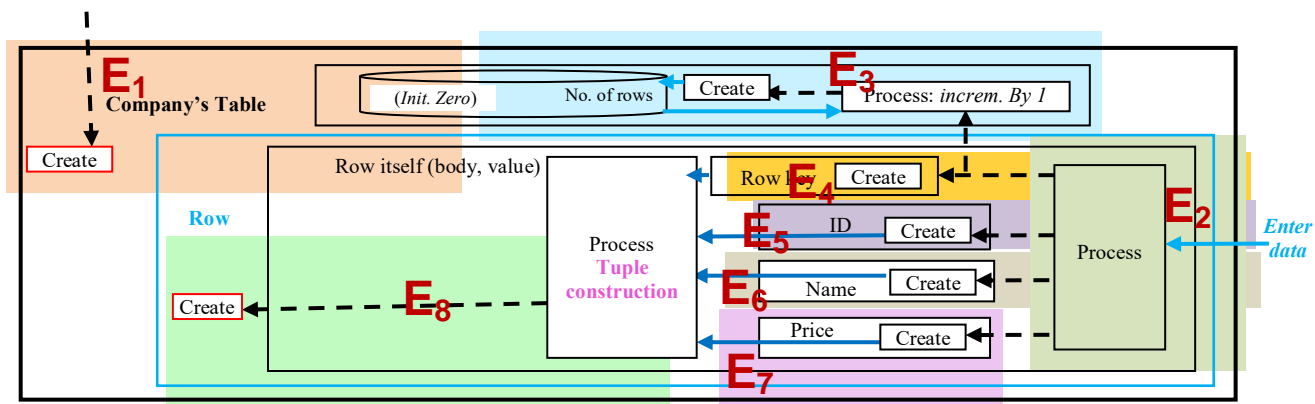

**Fig. 13 The events description of the example table**

This seems to be a promising approach to temporal databases. Note that in TM modeling, the stoppage of an event at a certain region is a return to staticity in that region [27]. Thus, TM modeling treats the absence (inexistence) of events as the subsistence of regions.

## IV. EXPLORATION OF THIMACS

Henri Bergson [28] distinguished two ways of knowing a thing: "The first implies that we move round the object; the second that we enter into it." This section provides a foundation for analyzing the *interiority* of the thimac as a machine (see Fig. 15). The purpose is to further understand the action *Create* that embeds the metaphysical meaning of the notion of existence in the next section.

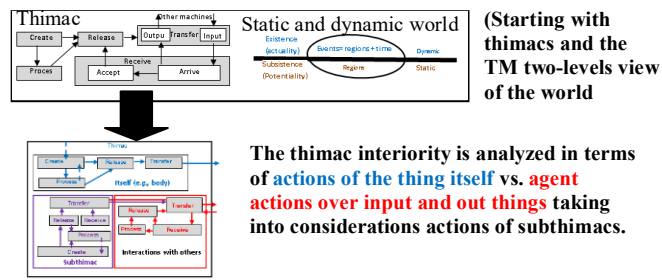

(Starting with thimacs and the TM two-levels view of the world

The thimac interiority is analyzed in terms of actions of the thing itself vs. agent actions over input and out things taking into considerations actions of subthimacs.

**Fig. 15 Content of section 4.**

### A. Two levels of TM: Potentiality regions vs. Actuality events

TM modeling describes reality in two levels of a potentiality/actuality scheme adopting an idea that goes back to the Stoic modes of reality. Fig. 16 defines the categorical structure of TM modeling. The two-level depiction is made to emphasize and illustrate the characteristics of each of the two levels; however, the two projected levels are superimposed over each other in TM modeling. Therefore, existence and subsistence are like a double-image impression (e.g., Rubin's vase), which is possible with a geostatic figure-ground perception. When we see an event, we simultaneously perceive its region. Many thinkers have inspired the general idea of this reality. For example, according to [29],

> The virtual [potential in TM] is not opposed to the real but to the actual. The virtual is fully real in so far as it is virtual [potential].... Indeed, the virtual must be defined as strictly a part of the real [actual] object – as though the object had one part of itself [TM region] in the virtual into which it is plunged as though into an objective dimension.... The reality of the virtual consists of the differential elements and relations along with the singular points which correspond to them. The reality of the virtual is structure.

### B. Regions and Events

A *region* is a TM-static description as illustrated in Fig. 17 in terms of the famous Aristotelian example of *statue ≡ wood + form*. The region has real subsistence, but such a type of reality is "absently present" [30]. The mind can conceive quasi-real subsisting things purely in itself without considering their existence, which is different from nonexistence (remember Rubin's vase).

An *event* in Fig. 17 is defined in terms of region and time. The event may denote the object (e.g., tree) or a process (e.g., traffic). Note that *mentality* is excluded in this TM discussion at this point of development of the model.

### C. Potentiality and Actuality

TM potentiality and actuality in Fig. 17 are different notions than Aristotelian's notions. Things subsist at the static level (plane of reality) and are brought above the threshold of subsistence/existence to actuality. In TMs, potentiality refers to all things at the static level that are *capable of existing*, i.e., all possible relative configurations. Events stand out against regions, and regions are a precondition to events.

For example, in the Aristotelian scheme, the piece of wood and the actual, say, *Hermes* statue are two ways (potential and actual) of being Hermes [31]. Then, potential Hermes and actual Hermes are two ways of being Hermes. In a TM, the piece of wood, form of Hermes and Hermes statue are three potential things that can be transformed into existence as illustrated in Fig 17.

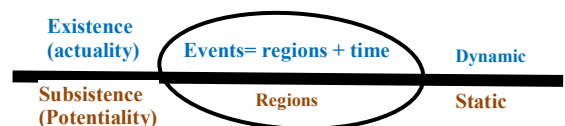

**Fig. 16 Two-level TM modeling**

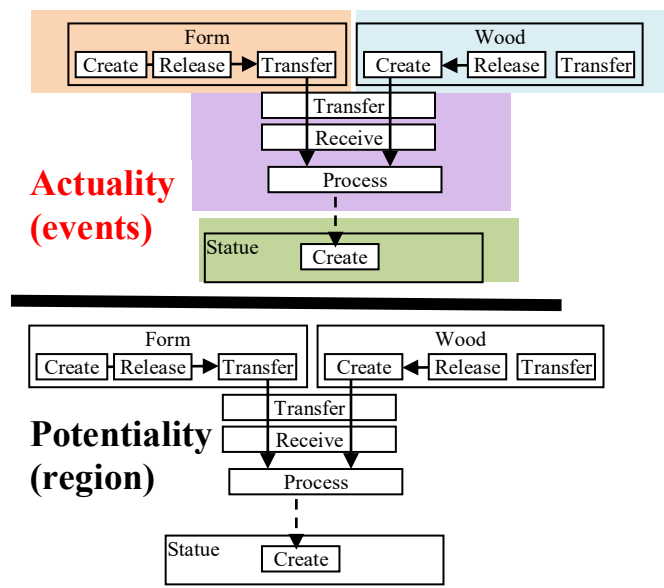

**Fig. 17 TM Potentiality and actuality for the Aristotelian example**

In the figure, wood, form and statue and processing them are a potentiality in the subsistence level that actualizes at the existence level.

The static level where time is suspended includes all chains of potentiality that unfold when time flows through regions, e.g., the event *man* (being one and the same) may materialize with a subthimac 'Pale' or in another time with a subthimac 'Dark,' just like an LED sign with *parts* turning the light on/off. Here, the fundamental aspect of change comprises various subdiagrams of the region. In the static level, regions are just as real as in present. According to [32], Abraham Lincoln is just as real as Joe Biden, "just as Venus is just as real as Earth: Lincoln is merely temporally 'far away from us,' just as Venus is spatially far away."

The potential thing would emerge into existence according to its time. The possibility that follows such an event is predetermined at the static level. Note that the arrows used in the static thimac are atemporal, e.g., in the classical example, the hand that moves the key in the lock of a door: the movement of the hand and that of the key are simultaneous, and yet one is prior to the other [33].

### D. Regions constituents

Region constituents define *what the thing* [33] (will be discussed in the next subsection) is or the being of a thing in the world. This definition is distinguished from the affirmation of the thing's existence (*the fact that it is* [33]). An existing thing is *composed* of a region and an *existence container* (will be defined later). Regions are a kind of *pre-existence* that freezes all actions and is the resemblance of events. A region is general in the sense that more than one event may have the same region (different locations in time). To preserve identity, it is not possible to have events of the same region and same location in time simultaneously. Regions and events are modes in which things come. When two events share the same region, then they are the same kind of event.

The definition of a region reflects a description of *all* potential permitted forms. A region has wholeness or an internal unity that is more than the totality of its constituents. If you have a region of a square and a region of a circle at the static level, then it is not possible to *create* a thimac of 'square circle' using elementary quantum particles/waves.

### E. Related classical notions: Nomads, Essence

Subsisting things in reality originate from first occurrences in reality [34]. For example, cars were not a subsisting thing until January 29, 1886, when Carl Benz applied for a patent for his vehicle powered by a gas engine. Before that, there were gas engines, cooling fans, etc. that fitted together to create the first car. After that, cars became subsisting things in the catalog of reality, where a new region is preserved in its own instances of existence. Thus, regions are not *found* independently of events, just as processes cannot exist without their events, e.g., traffic embedded into existing cars, roads, lights, etc.

Such a notion that a first thimac continues in the catalog of reality may be traced to *nature's* preservation of constituent species. A *natural thimac's* (e.g., tree) region subsists in miniature in the original thimac to have its own existence. The natural machine survives in its region; it can never be completely destroyed (see Leibniz's ideas about natural machines [35]). Each natural machine comes equipped with a substantial form (i.e., an entelechy that makes actual what is otherwise merely potential). According to Leibniz, the monad (as an existing thing) is made up of the organic machine (TM: the extension) in which innumerable subordinate monads come together that the dominating monad makes into one machine (TM: region) [35].

Following Avicenna, *to be a given entity in the world* (the fact that [something] is established/knowable/predicated) *without affirming its existence* must be distinguished from the existence of something. In establishing the "thingness" (essence) [TM region] of the thing, no existential judgment is implied, and it is independent of its existence. Such a distinction has an Aristotelian origin in which a thing exists separate from *what* a thing is. In TMs, this separation of thingness and existence is projected in the TM region at the static level and existence at the dynamic level.

Following the Aristotelian philosophy, *one thing* and *a thing* are the same thing. Now if a particular thing goes out of existence, it disappears from existence, but not from subsistence. In previous studies of thimacs, for the purpose of limiting study, the focus is only on thimacs mapped to existence.

## V.    THE THING ITSELF VS. OTHER THINGS

According to [5], "Metaphysical system building is model-building." This systematic metaphysics is work intended to be about the world itself and how the world is really constituted. *Model* is a hypothetical structure that we describe and investigate to understand some more complex, real-world target system or domain [5]. Specifically, in this paper, a model is understood as a high-level conceptual model in software engineering using diagrammatic methods.

In TM modeling, the thimac is the most important concept upon which the whole conceptual model of the notion of thimac is based. This thimac can be analyzed through two directions.

- The thimac itself is the *agent* that acts upon itself. For example, a human can move him/herself (*release/transfer*) from one place to another or *process* him/herself as in self-learning.

- The thimac is the *agent* that acts upon other things (objects). For example, when I experience things like cars, chairs, etc. that are not me.

Accordingly, the thimac involves different flows among actions: flows for itself, in its subthimacs and exterior thimacs that flow in/to the outside, as shown in Fig. 18 with blue, purple and red arrows, respectively. The red arrows indicate flows of other thimacs coming in from the outside, (maybe) processing and then leaving to the outside. There is no *create* in this type of flow because *creating another thimac* is a **triggering** mechanism to generate another thing. Thus, the creation is realized in the box of another thimac. The blue arrows involve the thimac itself and apply to the *whole* thing.

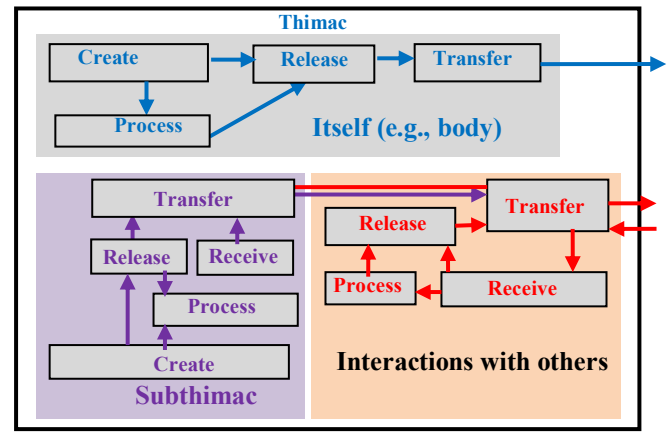

**Fig. 18 Thinging machine with different types of flow**

Fig. 19 shows how these three types of flows of TM merge together to eliminate redundant actions forming the TM machine, given previously in Section 2 as Fig. 4.

### A. Thing-in-itself

The TM thimac-in-itself, in which the thing as a whole is the object of actions, can be illustrated using the first example in Section 3 when *John Doe moves, then he moves in his totality*, i.e., body, *information about his father*'s reporting his birth, as shown in Fig. 20. The diagram does not reflect this movement explicitly; e.g., Father's data does not seem to move with John. This implicit information can be extracted by processing (extraction information) from John.

Such a distinction between the flow routes inside a thimac uncovers two meanings of the *Create* action as discussed in the next subsection.

### B. Create: transformation vs. persistence

Consider the TM stages from subsistence to existence levels that involve generating an event from a region. The thing is subject to alteration from the state of *static/dynamic transformation* to the state of *steady existence*. Certainly, the transformation process involves (1) the destruction of the region into pieces of generic actions; (2) giving each piece its substance (energy, materiality and so forth) then (3) constructing the region from these newly formed pieces into a region of an event. After this transformation to the existence level, the region would never be equivalent to the static region just as a house blueprint is different from the actual house.

Accordingly, there are two meanings of the *Create* action in TM.

- Create as *becoming* and
- Create as an *exis*tence *con*tainer (acronymized as **exicon**) that holds a thing (See Fig. 21). The term *existence container* is borrowed from [36] who stated, "The existence environment is an empty container much like a bottle. When a bottle is filled with milk, soda, juice, etc., it is referred to by the content of the bottle. You go to the refrigerator for milk, soda or a juice never thinking about the container, be it a bottle or can. Similarly, existence containers take on the properties of what is loaded into them, much like the bottle."

Note that the thing is a region (subdiagram) at the subsistence level; hence, it can be an *object* (e.g., tree) or a *process* (e.g. traffic).

*Becoming Create* (obtains or is given existence – an act of bringing into existence) that triggers (causes) the derivation of a new thimac, e.g., an *order* thimac triggers the creation of an *invoice* thimac. This Create is the source of an internal spark that leads to emergence from subsistence to existence and involves constraints, such as bodies, situations, etc. See this *becoming* Create in Fig. 22 (red letters).

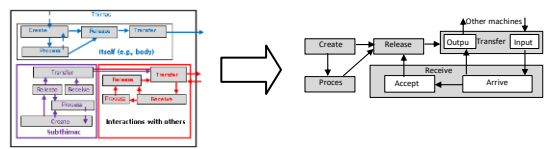

Fig. 19 Redundant actions (left) are eliminated (right).

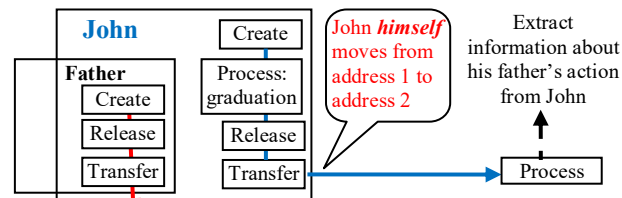

**Fig. 20 John Doe moves in himself.**

*Existence Create* preserves existence (persisting through time – see Fig. 22, green letters) after becoming. It represents the thing existence progressively in time *changing* (while existing) through actions: process, release, transfer and receive. For example, the famous ship of Theseus has this existence while loosening subthimac parts (un-manifestation at the static level) and gaining parts (manifestation at the dynamic level) through the history of its repairs. We conceive the ship at the static level as a pure placeholder to which various parts attached.

### C. Existence Container

In this section, the focus is on *Create* as *persisting existence* in an attempt to isolate existence in its pure form: **exicon**. Consider Fig. 23, which shows the events,

*John exists*, and

*John creates an order and sends it somewhere*.

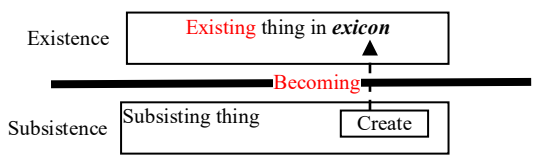

**Fig. 21 The role of *Create* that transforms a region into event**

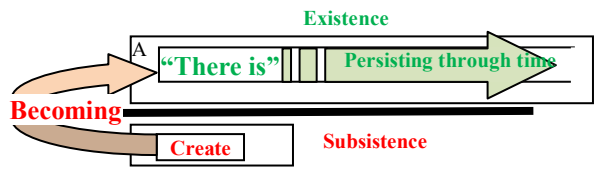

**Fog. 22 TM *becoming* and persisting**

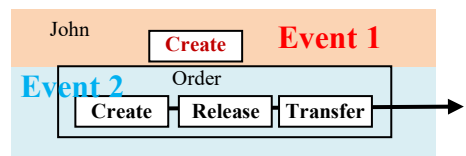

Fig. 23 Two events: *John exists and sends an order*

Let us concentrate on the event *John exists* (Fig. 24). *Create* in TM is different from the other four TM actions in given a region. It stands for the existence of a thimac. After becoming, the *create role* in the region of an event alters to *there is* (see Fig. 25). At the dynamic level (existence), *create* stands for *there is*. Here, we can define *existence* as,

**Existence of a thing ≡ There is a thing in time**

Consequently, the TM diagram can be simplified as *John exists* as shown in Fig. 26.

Suppose that we define *pure* existence (existence without specific thimac or time), previously called an *exicon*, as the blend of time and other phenomena (e.g., entropy [decay] and energy) that regions soaked in (see Fig. 27). This exicon is born (becoming) to go through decay and fade away when it becomes part of the past. It encompasses time and change aggregated according to events.

The exicon has no specific event. It may be similar to *pure awareness* without thought or attention. Awareness is different from the objects of awareness. Also, the exicon is existence without event. The exicon embraces things just as awareness of ideas/concepts is the content of thought. Things stand up in existence and fade away from existence, while existence itself seems to continue unaffected.

Future research can explore the relationship of this pure existence to Leibniz's monad as constituents of the world. The word *monad* is said to be used by Pythagoras around 500 BCE to refer to the first being that came into existence: the *absolute source of creation*.

An exicon can be described as a 'slot' of existence (of different sizes). John Dewey [37], rejecting the classical metaphysics of substance, stated that "Every existence is an event." In TM, every event is an existence that occupies an exicon.

### D. Glimpse of related metaphysical issues

An exicon can be described as a slot of existence, say, *C* to be filled with an event, *e*, thus it can be specified as *C(e)*. According to [38], "*Is there* any such-and-such?" (Emphasis added) means merely, "Restricting our attention for the moment to just things that are so-and-sos, is there a such-and-such among them?" hence, in a party,

When we ask, for example, "Is there any beer?," we usually mean merely "Restricting our attention to just beverages in the fridge, is there any beer?" If the last beer has been taken from the fridge at a party, and someone asks, "Is there any beer?," it is a poor joke to say "Yes" and then explain that there is plenty in the grocery store (which is closed, by the way). The metaphysician interested in ontology wants to know what the world is like in its entirety, ignoring nothing.

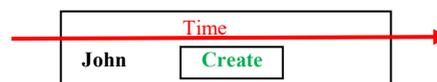
**Fig. 24 Event: John as an object**

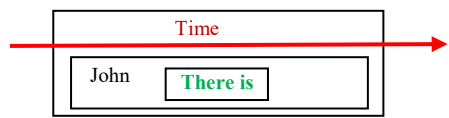
**Fig. 25 Event: *John as an object***

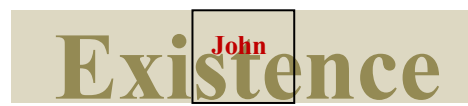
**Fig. 26 The *John* in existence**

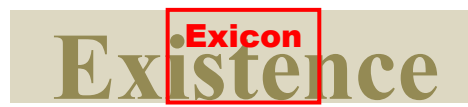
**Fig. 27 Capsule of existence**

In TM, "Is there any beer?" means *C(beer)?* While the correct question is does *C(refrigerator)* include *C(beer)*? Note that all actors: the questioner, the one who replies with "yes," the refrigerator and the beer are exicons. Thus, an exicon asks about an exicon in the exicon to be answered by yet another exicon.

This may be called a *democracy of exicons*, a generalization of the so-called democracy of objects approach that adopts the thesis that "all objects ought to be treated equally. The democracy of objects is the *ontological* thesis that all objects equally exist while they do not exist equally. The claim that all objects equally exist is the claim that no object can be treated as constructed by another object" [39].

### VI. CONCLUSION

This paper has introduced an exploration of metaphysical origins of conceptual modeling as exemplified by a specific proposed high-level model called thinging machines (TM). The results showed several interesting results about the nature of thimacs at the static and existence levels. Especially, the analysis has led to define existence in TM modeling. The topic is too extensive and requires further exploration including,
- Related notions such as Leibniz's Monads and Avicenna's Essence.
- Related existence notions such as the democracy of exicons.

It is interesting that research that started with conceptual modeling in software engineering has headed to very deep issues in metaphysics. The interdisciplinary ramification of such an approach requires further research.